\documentclass{optica-article}

\journal{opticajournal} 

\articletype{Research Article}

\usepackage{lineno}
\nolinenumbers 

\begin{document}

\title{Optical switching beyond a million cycles of low-loss phase change material Sb$_2$Se$_3$.}

\author{Daniel Lawson\authormark{1,2}, Sophie Blundell\authormark{1,2},  Martin Ebert\authormark{2}, Otto L. Muskens\authormark{1,2}, and Ioannis Zeimpekis\authormark{2,*}}

\address{\authormark{1}Physics and Astronomy, University of Southampton, Southampton, UK\\
\authormark{2}Optoelectronics Research Centre, University of Southampton, UK}

\email{\authormark{*}izk@soton.ac.uk}


\begin{abstract*} 
The development of the next generation of optical phase change technologies for integrated photonic and free-space platforms relies on the availability of materials that can be switched repeatedly over large volumes and with low optical losses. In recent years, the antimony-based chalcogenide phase-change material Sb$_2$Se$_3$ has been identified as particularly promising for a number of applications owing to good optical transparency in the near-infrared part of the spectrum and a high refractive index close to silicon. The crystallization temperature of Sb$_2$Se$_3$ of around 460 K allows switching to be achieved at moderate energies using optical or electrical control signals while providing sufficient data retention time for non-volatile storage. Here, we investigate the parameter space for optical switching of films of Sb$_2$Se$_3$ for a range of film thicknesses relevant for optical applications. By identifying optimal switching conditions, we demonstrate endurance of up to 10$^7$ cycles at reversible switching rates of 20~kHz. Our work demonstrates that the combination of intrinsic film parameters with pumping conditions is particularly critical for achieving high endurance in optical phase change applications.
\end{abstract*}

\section{Introduction}
Phase change materials (PCMs) have the ability to undergo a non-volatile transition between amorphous (glassy) and crystalline states which possess different optical, electrical and thermal properties \cite{Ovshinsky1968ReversibleStructures,Wuttig2007Phase-changeStorage, Raoux2009AnnRevMat, Burr2016,Wuttig2017}. Their historical development is tightly wound into the history of computing. During his lecture to the London Mathematical Society in 1947, Alan Turing advocated the need for a practical form of memory to replace the existing punch cards \cite{Turing2004Lecture1947} and subsequent technologies were developed. The first integrated circuits were invented by Jack Kilby of Texas Instruments \cite{Kilby}, opening the door to solid state memory and random-access memory (RAM). In 1967, D. J. Shanefield demonstrated the first phase change memory device, featuring electrical switching \cite{Shanefeld1966OperatingDevices}. 
This followed on from discoveries of the switching characteristics of phase change materials in the early 1960s \cite{Pearson1962ChemicalGlasses}, followed by Stanford Ovshinsky's well-known work \cite{Ovshinsky1968ReversibleStructures}, demonstrating what would later be named ``Ovshinsky threshold switching'' (OTS). The main feature of this switching, commonly observed in chalcogenide PCMs, is that once a crystalline material has reached the threshold, it rapidly transforms to become amorphous. 
Ovshinsky measured this as a conductive (crystalline) and resistive (amorphous) state.

Further research into how crystallization speeds are controlled, such as choice and composition of PCM, gave gradual increases in rapidity of switching, particularly through the developments and advancements in the switching of the phase change material GST\cite{Yamada1991RapidphaseMemory}; in 1986 GeTe was reversibly switched with 100~ns pulses \cite{Chen1986CompoundStorage}, and by 2012 500~ps switching of GST was possible \cite{Loke2012BreakingMemory}.  Not long after in 2010, two independent works by Konishi and Fons and their co-authors evidenced the first cases of sub-picosecond switching in PCMs \cite{Konishi2010, Fons2010}. \

For many years PCMs have been also successfully employed as electronic and optical storage materials \cite{Yamada1991RapidphaseMemory, Burr2016,Raoux2009AnnRevMat}. In the 1990s, Panasonic developed phase change materials for optical data storage, releasing their first optical disc, which featured Sb\textsubscript{2}Te\textsubscript{3}-GeTe \cite{Yamada1991RapidphaseMemory}.  Crystallization of amorphous PCM and amorphization of crystalline PCM can be performed by controlling electrical current, which invokes Joule heating \cite{Lankhorst2005Low-costChips}. Optical switching of PCMs uses laser irradiation, rather than electrical pulses, to heat the material which allows the material to reach both the crystallization and melting temperatures when required. 

The interest of using PCMs in photonics applications has seen a rapid growth in recent years owing to the increasing maturity of the fields of integrated photonics, metamaterials, photonic computing and AI where programmable and switchable functionalities are extremely desirable \cite{Wuttig2017,Abdollahramezani2020TunableMaterials,Miller2018OpticalReview,Li2023NeuroReview,Zhou2022MRSBull,Rios2021Ultra-compactMaterials}. However photonics applications are lagging behind their electronic counterparts, mainly due to the lack of specific design of PCMs with favourable optical properties. This is partly due to the tailoring of PCM properties for electrical memory applications, where high switching speeds and durabilities were optimized specifically for electrical switching configurations \cite{Gholipour2013AnMeta-Switch}. 
In recent years, increasing efforts have been directed towards the exploration of previously unstudied families of PCMs. Examples include the extension of the Ge-Sb-Te family, which yielded Ge$_2$Sb$_2$Se$_4$Te$_1$ (GSST) as an optimized material for optical applications \cite{Zhang2019}. Also, PCMs that show a dielectric-to-metal transition such as In$_3$Sb$_2$Te$_2$ have been considered as particularly interesting for devices exploiting plasmonic effects, with especially high contrast in the terahertz region \cite{Heßler2021,Heßler2022,Zeng2023}. A class of PCMs of particular interest for low-loss and high-refractive index applications in the near-infrared are the antimony-based binary chalcogenides Sb$_2$S$_3$ \cite{Dong2018} and Sb$_2$Se$_3$\cite{Delaney2020}. The requirements for optical PCMs are very different from their electronic counterparts, as has been critically discussed in several recent works \cite{Zhang2021,Simpson2}. The electro-optical-thermal response of the materials has to be carefully designed to achieve compatibility with specific applications\cite{ZhangMyths}.

In this work we explore two aspects in the integration of the PCM Sb$_2$Se$_3$ for reconfigurable photonic applications, namely the durability and the optical switching of this material when deposited in films of thickness from 80~nm up to 800~nm. This extended range of PCM thicknesses is relevant for applications in free-space optical components such as nanophotonics, diffractive optics and metasurfaces. The parameter space of optical pumping is optimized through systematic studies of the amorphization and crystallization conditions. We furthermore identify that, for films with thickness exceeding 200~nm, the optical pumping conditions result in a spatial energy distribution near the top of the layer and therefore a partial amorphization.  We note that this behaviour is specific to our choice of pump laser at a wavelength of 480~nm, where the skin depth is only a few tens of nanometers. While fairly narrow windows of operation are identified, these allow selection of optimized pulse parameters for the reversible cycling of the PCM film at high cycling rates of up to 20~kHz when operating on a crystalline background for film thicknesses below 150~nm. The choice of optimized pumping conditions allow reversible switching over more than 10$^6$ switching cycles. The demonstration of such high endurance increases the range of potential applications within integrated photonics and further into free-space reconfigurable PCM devices. In such applications there is a potential need for reversible switching of PCM components with thicknesses on the order of many hundreds of nanometers.

\section{Background}

To crystallize an amorphous PCM, the generally accepted model is that of ``steady state crystal nucleation'', which was originally developed by Gibbs \cite{Gibbs1876OnSubstances}, Volmer \cite{Volmer1939KinetikPhasenbildung}, Weber \cite{Volmer1926KeimbildungGebilden}, Becker, D{\"o}ring \cite{Becker1935Annal}, Turnbull and Fisher\cite{Turnbull1949RateSystems}. 
This describes how crystallization is initiated by crystal nucleation, which is followed by a stage of crystal cluster growth, once a stable crystal cluster has developed. 
A crystal cluster is energetically favourable when the cluster grows larger than a critical radius.
This is an activation barrier that allows a liquid to be cooled without crystallization occurring. 
It has been found that Ge-Sb-Te (GST) alloys typically exhibit nucleation-dominated crystallization while Sb-rich alloys tend towards growth-dominated crystallization \cite{Salinga2013, Rao2017ReducingWriting,Muller2022, Wang2023npj}. 
Practically, crystallization is incurred by heating the material above its crystallization temperature and allowing time, at that temperature, for nucleation and growth to occur. 
Crystallization by growth from crystalline-amorphous boundaries can occur as well as from the nucleation centre outwards \cite{Meinders2006OpticalRecording}.

Amorphization generally requires the material to be heated above its melting temperature, then rapidly cooled in order to ``fix'' this seemingly random arrangement of atoms. 
This rapid cooling should be of a rate on the order of 10\textsuperscript{9}~Ks\textsuperscript{-1} for poor glass formers such as GST \cite{Wuttig2017}.
Many materials can be melt-quenched in order to form an amorphous state (such as silicon), but the novelty of PCM glasses are the vast differences in optical properties between these two states \cite{Wuttig2007Phase-changeStorage} and the accessibility of stable amorphous phases with moderate cooling rates. In some chalcogenide PCMs such as GeTe slow crystallisation can be observed, on the order of tens of microseconds, which can alleviate the necessity for high cooling rates\cite{Raoux2009}.

Many studies have been performed which investigate existing PCM materials, ranging from doping strategies to heterostructure design. By combining ultra-thin layers of different PCMs, heterostructure geometries can be designed which inherit the optical properties of their parent materials and exhibit some benefits over their bulk counterparts\cite{Meng2023het,Ding2019het,Zeng2023het,Kalikka2016het,Simpson2010,Zhou1999}. Interfacial effects in such structures can lead to drastic reduction in the energy threshold for reset operation. In the case of the TiTe\textsubscript{2}/Sb\textsubscript{2}Te\textsubscript{3} heterostructure devices demonstrated by Ding et al. \cite{Ding2019het} a substantial improvement in the amorphous phase stability and repeatability can be seen. Doping can be used to drastically alter the crystallization of PCMs, with metallic dopants generally increasing the speed of crystallization. Furthermore, the careful design of the PCM and surrounding geometries can be used to control the thermal transport properties of PCM components. The choice of capping layer and interface materials for example, can have a considerable impact on the crystallisation dynamics.\cite{Teo2023, Burr2016}

Traditional GST electronic PCMs integrated into nanoscale mushroom cells demonstrate a very large cycling durability, with property contrasts maintained even up to 10$^{12}$ cycles between the crystal and glass states \cite{Kim2017}. In comparison, demonstration of photonic devices have shown much lower cycling durabilities and yet the optimization of the durability performance has remained largely unexplored with most experimental demonstrations claiming between 10-1000 cycles \cite{Martin-Monier:22, Fang2022Ultra-lowHeaters}.

Gholipour et al. \cite{Gholipour2013AnMeta-Switch} reported on switching of GST photonic devices limited to $<$ 50 cycles and recommended exploration of new families of materials to find a more robust substitute.
It has since been shown that GST, in certain configurations, is capable of  >1000 cycles before degradation \cite{Zheng2020NonvolatileHeater}. 
However, since GST absorbs at telecommunications \cite{Zhang2017BroadbandMaterials} as well as at visible wavelengths  \cite{Fang2021Non-VolatileMaterial}, reportedly has a bandgap of $\sim$0.5~eV \cite{Lee2005InvestigationPhases}, low thermal conductivity \cite{Zhang2019}, requiring a high cooling rate and film thickness $<$ 100~nm, the search for a new family of materials has only become more pressing.

A recent replacement has been GSST, which is transparent in the visible and near infrared (NIR) ranges (1-18.5~$\mu$m), can be switched in larger volumes than GST, has a refractive index change, $\Delta$n, of $\sim$2 at 1550~nm (with refractive index  of  3.38 + 0.00018i and 5 + 0.42i for the amorphous and crystalline phases respectively) \cite{Zhang2018OL,Zhang2019,Xu2022}. GSST also displays moderate endurance, with $>$1000 cycles before degradation \cite{Lepeshov2021TunableMetasurfaces,Zhang2019,Xu2022}.
The reduced losses of GSST in the near-infrared are associated with a smaller optical contrast between the two phases compared to GST. Also reported, is the reduced thermal conductivity in GSST, attributed to a switch in thermal transport mechanism from using electrons to phonons when switching between phases \cite{Aryana2021SuppressedGe2Sb2Se4Te}. 

GST-related alloys are often considered, or compounds of its components, such as Sb$_2$Te$_3$, which operates in the infrared regime and reportedly has $\Delta$n $>$ 2 \cite{Moon2019ReconfigurableMaterials}.
Sb$_2$Te$_3$ can be improved by doping with titanium.
This improves thermal stability, reduces the change in density to just 1.37\%, reduces the grain size in the crystalline phase from 200~nm to 12~nm, which helps with adhesion to a substrate, and maintains a switching time of ~10~ns \cite{Xia2015Ti-Sb-TeMemory}.
TiSbTe does however require a high operational voltage, which is attributed to doping of excessive Ti.

In the last decade, interest in antimonides has increased, with Sb$_2$S$_3$ and Sb$_2$Se$_3$ becoming popular PCMs for photonic applications \cite{Tang2017AntimonyFibers, Mamta2021ASelenide, Delaney2021NonvolatileMaterial,Wang2022ReconfigurableSb2Se3, Lei2022Magnetron-sputteredPhotonics}\cite{Teo:22}. 
Sb$_2$S$_3$ has a bandgap in the visible regime of 2-1.7~eV, a crystallization temperature of $>$573~K, a maximum $\Delta$n of 1 at 614~nm, and a switching time of 70~ns \cite{Dong2019WidePhotonics}. 
It performs well within visible wavelengths, showing a 30~dB modulation contrast \cite{Fang2021Non-VolatileMaterial} but does experience a 35~\% increase in density upon crystallization \cite{Dong2019WidePhotonics}.
Sb$_2$Se$_3$, in comparison, works better in the telecommunication C-band (1550~nm), experiencing very low losses (of k $<$ 10\textsuperscript{-5}) \cite{Delaney2020}.

\subsection{Switching of thicker PCM layers}
In the pursuit for optically reconfigurable metasurfaces comprising  purely PCM elements on a substrate, one might consider individual PCM nanopillars which serve as the individual pixels of PCMs, or PCM metaunits \cite{Shalaginov2021}. For a metasurface to function as an arbitrary optical phase modulator, each PCM metaunit must be able to induce a full 2$\pi$ phase shift when switched from an amorphous state to a crystalline one. Then, each unit may be programmed to different levels of crystallinity, as to be able to select the local optical phase at each pillar by manipulating the refractive index change between states. This demonstration, however, is fundamentally limited by the capacity of common PCMs to fully transition between the amorphous and crystalline states when the active volume becomes large and especially along the direction normal to the plane. This limitation is commonly known as the thickness limit.

A common explanation for the origin of the thickness limit in phase change materials follows the relationship between the intrinsic crystallisation properties, namely the minimum time for crystallisation $\tau_{\rm min}$  and the thermal conductivity $\alpha$, and the quenching rates of melt-quenching glass. In an ideal amorphisation process, a superheated liquid-phase volume is generated by heating the PCM above its melting point T\textsubscript{m}. The PCM is then allowed to cool below the nose temperature T\textsubscript{nose}. If the total temperature change required is dT\textsubscript{mg}, $dT_{\rm mg} = T_{\rm m} - T_{\rm nose}$ then the quenching rate can be written as $dT_{\rm mg}/t$. If $dT_{\rm mg}/t$ is sufficiently high, then the PCM can vitrify, forming an amorphous glassy state with a lack of medium-range order. For the majority of PCMs used in reconfigurable optics, $dT_{\rm mg}$ is on the order of several hundred Kelvin, and is the maximum time before the onset of crystal growth. For example, GST is known to be able to crystallise on the order of a few nanoseconds. The critical quenching rate $\Theta_{\rm crit}$ is the cooling rate required for $\tau_{\rm min}$. Here $\tau_{\rm min}$ refers to the minimum time required for the onset of crystallisation in the quenching of the amorphous phase $dT_{\rm mg}$ and $\tau_{\rm min}$ requires $\Theta_{\rm crit} \sim 10^{9}$~K/s for GST-225 due to $\tau_{\rm min}$ values on the order of nanoseconds during a thermally-driven transition. For a full depth transition from a crystalline phase to an amorphous phase the thickness limit $t_{max}$ can be written as

\begin{equation}
\mathrm{t}_{\max } \approx \sqrt{\frac{a \cdot\left(\mathrm{T}_{\mathrm{m}}-\mathrm{T}_{\text {nose }}\right)}{\theta_{\text {crit }}}}
\end{equation}

assuming constant $\alpha$ throughout the film depth and a homogeneous critical rate $\Theta_{\rm crit}$ \cite{ZhangMyths,Zhang2021}. This critical thickness is a crucial limitation on the demonstration of meta-structures based on meta-atoms formed solely out of PCMs. Reflective geometries indeed have been shown to reduce the height requirements for 2$\pi$ phase control. Typically in materials with $\tau_{\rm min}$ < 100~ns (i.e GST, Sb$_2$Te$_3$ and Sb$_2$S$_3$)  other dielectric or plasmonic components are utilised instead as optical resonators, whilst using a surrounding PCM as an active layer for the modification of the resonance behaviour\cite{Gholipour2013AnMeta-Switch,Wang2021,Tittl2015,Michel2014,Zhu2022DynamicallyChalcogenides,Abdollahramezani2022}. It is then possible to reduce the thickness requirement drastically, thus enabling reversible switching via optical, electrical and thermal routes. Wang and co-authors demonstrated recently both a tunable varifocal and intensity modulating metalens designs based on Sb$_2$Se$_3$ nanopillars, operating at a wavelength of 1550~nm \cite{Wang2023}. With careful selection of the pillar dimensions to utilise electric and magnetic dipole resonances 2$\pi$ phase control is achieved. The thickness of the pillars required for operation at 1550~nm is 270~nm. The thickness limit $t_{max}$  varies with the material's specific heat capacity and thermal conductivity but also with the minimum crystallisation time. Considering $\tau_{\rm min}$  values for Sb$_2$Se$_3$ around 800~ns, as demonstrated in this work, we can therefore derive that similar to GSST \cite{Zhang2021},  Sb$_2$Se$_3$ can be switched for up to at least 1~$\mu$m of thickness, far exceeding the requirements demonstrated by Wang et al. Herein we focus on the challenge of dynamic and reversible switching of PCMs with thicknesses below the stated $t_{max}$ via pulse optical heating, which is particularly applicable to highly efficient meta-optic designs without the incorporation of lossy plasmonic components. 

\section{Methods}

\begin{figure}[h]
\centering\includegraphics[width=\textwidth]{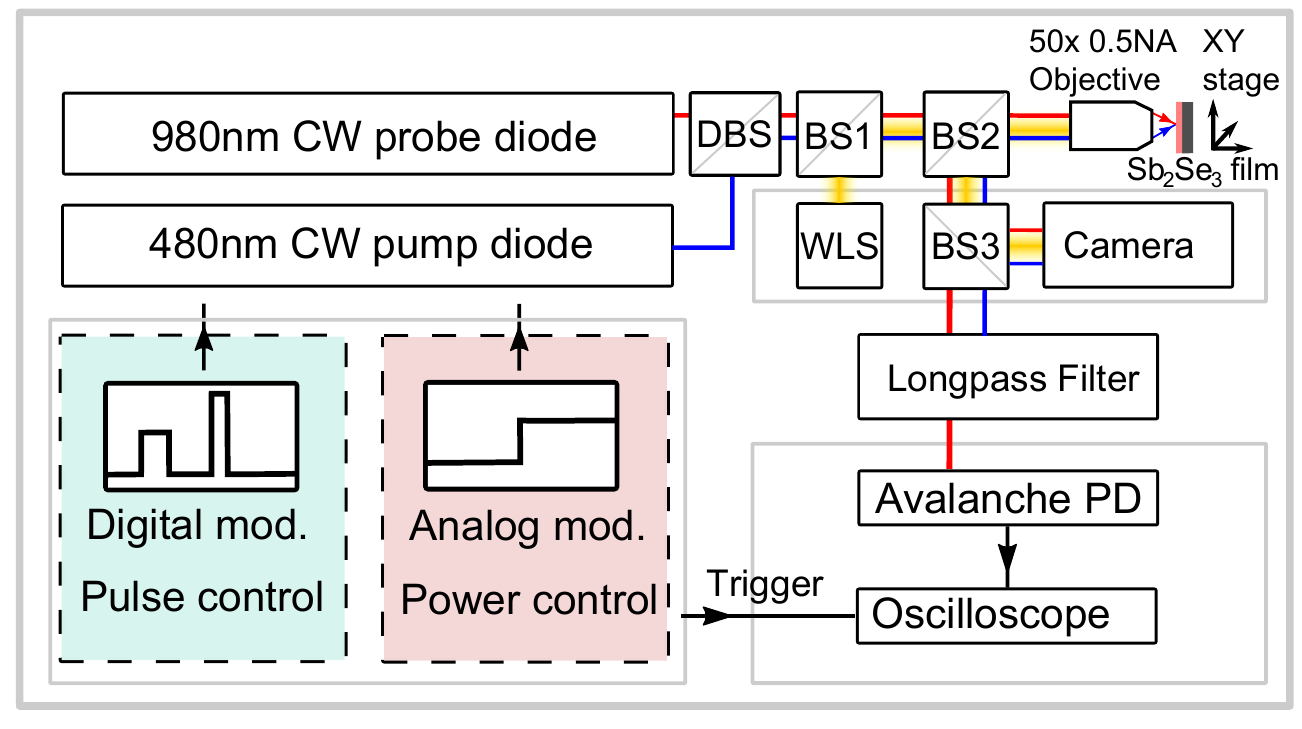}
\caption{Schematic diagram of experimental setup for optical switching and endurance testing of Sb$_2$Se$_3$. Setup consists of continuous wave (CW) current-modulated pump diode at 480~nm wavelength, digital and analog pulse generators, near-infrared CW probe laser at 980~nm, dichroic beam splitter (DBS), beam splitters (BS 1-3), white-light source (WLS), Motorized sample positioning stage, CMOS camera, near-infrared longpass filter, fast avalanche photodetector (0-400 MHz) and oscilloscope (Picoscope 5000).\cite{Lawson2022}}
\label{fig:setup}
\end{figure}

In this work, the switching behaviour in films of thermally-annealed crystalline Sb$_2$Se$_3$ films was characterized by {\em in-situ} measurements of the reflectance change of a 980~nm continuous wave (CW) probe laser in a static-tester station \cite{Behera:17}. Figure~\ref{fig:setup} shows a schematic overview of the experimental arrangement. The probe laser wavelength of 980~nm was chosen in order to minimize heating of the Sb$_2$Se$_3$ layer during relflectance measurements. A current-modulated diode laser at operating wavelength of 480~nm was used to produce optical pulses of varying power and pulse duration to induce change in the PCM between its crystalline and amorphous phases. In the case of the optical static tester maps we use a arbitrary function generator (Berkeley Nucleonics) to set the pulse power and length via triggered analog and digital modulation signals controlled by a LabVIEW computer script. For the endurance measurements we use a free-running arbitrary function generator (B\&K Precision) to provide both the digital and analog modulation, allowing us to cycle the state of the Sb$_2$Se$_3$ films at cycling rates of up to 20~kHz. This approach allowed to perform the endurance measurements within practically achievable timescales over which the effects of optical drift of the setup are small (from 50~seconds for 10\textsuperscript{6} cycles up to 83~minutes for 10\textsuperscript{8} cycles at a cycling rate of 20~kHz). The thin-film sample was mounted on a motorized dual-axis translation stage with a manual three-axis mount for tilt and yaw control, which was used to eliminate any focus drifting which may arise due to  translation of the samples.

We found that primary sources of instrumentation error are caused by slow drifting of the microscope focus on a time scales of hours, which imposes an instrumental limitation in very long burn-in experiments. A small periodic oscillation is seen in some of the data at time scales of milliseconds, which can be attributed to spurious variations in laser intensity at characteristic frequencies below 200~Hz. A much slower periodic variation on time scale of minutes can be attributed to slow oscillations of the temperature in the room of around 0.2~$^\circ$C, due to an active feedback cycle in the air conditioning system. This very small temperature variation acts back onto the pump power through the extreme sensitivity on the focus of the microscope, which is the primary cause of the vertical bands seen in the static tester maps.

For the deposition of the Sb\textsubscript{2}Se\textsubscript{3} films we used an AJA Orion RF magnetron sputterer with a stoichiometric 3-inch Sb\textsubscript{2}Se\textsubscript{3} target. For all fabrication runs the main chamber pressure was fixed to 5.75~mTorr with an argon gas flow rate of 20~sccm, and 35 W RF power on the target. All films were deposited on p-type, 1-10 ohm cm, Si wafers with a [100] orientation. After deposition of the PCM films a ZnS:SiO\textsubscript{2} capping layer of 200~nm thickness was deposited without breaking vacuum. All samples were then annealed at a temperature of 200~$^\circ$C for 5 minutes to crystallise the deposited films.

\section{Results}

\subsection{Laser-induced amorphization of Sb$_2$Se$_3$}
\begin{figure}[ht!]
\centering\includegraphics[width=0.95\textwidth]{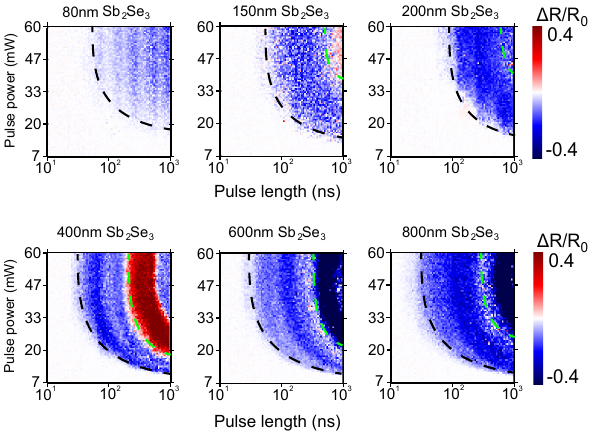}
\caption{Optical amorphization maps showing relative reflectance change $\Delta R/R$ for amorphous volumes written at 10 - 1000~ns pulse duration and 7 - 60~mW of average laser power. Maps were taken from a crystalline background obtained by annealing samples at 200$^\circ$C for 30 minutes. Dashed line, black: guide to the eye denoting switching threshold. Dashed line, green: guide to the eye denoting damage threshold.}
\label{fig:amo}
\end{figure}
The optical switching of Sb$_2$Se$_3$ films with thickness ranging from 80~nm to 800~nm was studied in order to characterize the effect of the layer thickness on the parameter window in which amorphization is obtained. Permanent damage is typically distinguished as failure to recrystallize the written regions with a low power, long duration pulse and typically appear as persistent dark spots in the sample when inspected under an optical microscope. We used this method to identify the range of amorphization pulse parameters resulting in effective switching below the damage threshold. 

To gauge the pulse regimes that give rise to effective switching or damage in the amorphization process, we performed parameter sweeps of the pump pulse length and power. Figure~\ref{fig:amo} shows the change in reflectance induced by the pump pulse normalized to the reflectance before the pulse, $\Delta R/R_0$. The black dashed lines in Fig. \ref{fig:amo} indicate the amorphization thresholds of the Sb$_2$Se$_3$ films. These lines are drawn as a guide to the eye, but practically follow where the amorphization produces a reflectance change greater than 1~\%. The general trend seen here with increasing Sb$_2$Se$_3$ thickness, is that the amorphization threshold decreases, with the switching threshold moving to lower pulse energies for films up to 400~nm thick. We attribute this change to a combination of overall increased absorption in the thicker films and a concomitant decrease in the vertical thermal transport as the film's own thermal resistance acts as a barrier to the silicon substrate's higher thermal conductivity. Together these effects allow for the formation of surface molten volumes at lower pulse energies in the thicker films. For films above 400~nm thickness, we see no change in the switching threshold.
\begin{figure}[ht!]
\centering\includegraphics[width=\textwidth]
{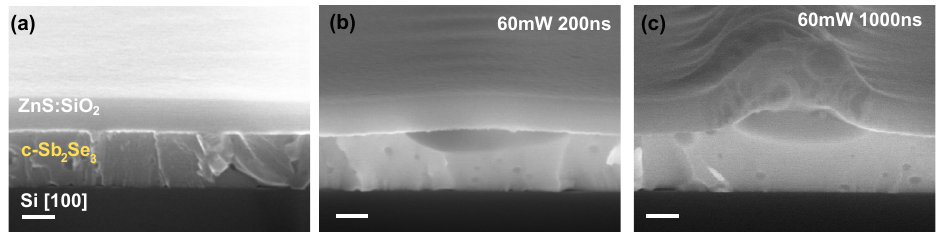}
\caption{Scanning electron microscopy (SEM) images of cross sections of the 400~nm Sb$_2$Se$_3$ film with 200~nm ZnS:SiO$_2$ cladding, after crystallization using hot plate (a), and for optically-drive amorphization at 60~mW laser power (b,c) and pulse durations of 200~ns (b) and 1000~ns (c) respectively. Scale bars in (a-c) are 200~nm.}
\label{fig:cross}
\end{figure}
Figure \ref{fig:amo} also shows the damage threshold of the amorphization as indicated by the green dashed lines. This damage threshold is seen as an abrupt change in the optical reflectance. This irreversible change in the reflectance could be attributed to a significant change in the capping layer morphology as shown in Fig.~\ref{fig:cross}(a-c). Here, we compare cross-sectional scanning electron microscopy (SEM) images of the 400~nm Sb$_2$Se$_3$/ZnS:SiO$_2$ stack without laser pulse (a), with amorphization using a pulse parameter of 60~mW and 200~ns (b), and for pulse parameter of  60~mW and 1000~ns (c). The condition shown in Fig.~\ref{fig:cross}(b) corresponds to the regime of reversible excitation and shows an amorphous top region of the Sb$_2$Se$_3$ film extending to about 200~nm into the layer. The expansion of the amorphous lattice gives rise to a bulging of the top surface which is accommodated by the 200~nm thick cladding layer indicating the need for a more substantial capping. For the case of optical pumping at 5 times longer pulse duration, we see a ballooning of the top layer, indicating loss of structural integrity of the stack which is irreversible. Although the loss of integrity of the top layer does not cause direct damage to the PCM volume itself, it is expected that this damage compromises the encapsulation of the Sb$_2$Se$_3$,  as a direct result of excessive heating and the volume change upon a crystalline to amorphous phase transition. Consequently this can result in oxidization of the film, and loss of the switching properties in following cycles. 

Figure~\ref{fig:cross} furthermore shows that for the 400~nm thick film, the amorphous region extends only 200~nm into the Sb$_2$Se$_3$ layer, thus indicating that for films beyond a thickness of a few hundred nanometers this method of optically induced switching does not result in a fully amorphized region. Clearly the extent of the amorphization volume is determined by the interplay between optical absorption profile, heat diffusion and quenching dynamics. In our studies, the optical absorption depth at 480~nm wavelength is likely to be the limiting factor determining the initial distribution of energy into the layer, with the 1/e absorption depth being 14.48~nm for a refractive index of Sb$_2$Se$_3$ of 4.41 + 2.64i in the crystalline state at the pump wavelength \cite{Delaney2020}.

\subsection{Laser-induced crystallization of Sb$_2$Se$_3$}

\begin{figure}[ht!]
\centering\includegraphics[width=0.95\textwidth]{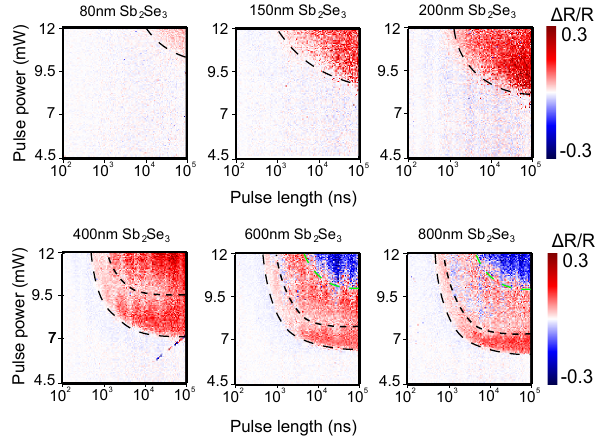}
\caption{Optical recrystallization maps for amorphous volumes written at 60~mW power and 100~ns pulse duration. The black dashed lines indicate the switching threshold and the transition between the ideal window for crystallization and a region of reduced reflectance contrast between the PCM states. Damage is seen in the regions above the green dashed lines. }
\label{fig:recrys}
\end{figure}

Having identified the parameter map for pulsed laser amorphization of the PCM, recrystallization maps of amorphous domains were produced in order to find the optical pulse regime required for a fully reversible transition back to a crystalline state. First, amorphous volumes were written using a fixed amorphization pulse power of 60~mW and duration of 100~ns, which was identified to be below the damage threshold for all films and within the range where optical switching could be achieved. Next we performed parameter sweeps of the pulses used in the recrystallization process from 100~ns to 100~$\mu$s. Our work distinguishes from previous studies\cite{Lawson2022}\cite{Muller2022} where crystallization dynamics were studied starting from a pristine amorphous material. In that case, crystallization times of milliseconds were reported for Sb$_2$Se$_3$, limited by nucleation dynamics. In contrast, in this work we operate at the regrowth of crystalline domains from a crystalline background surrounding a small amorphous region, which happens on microsecond time scales.

Figure \ref{fig:recrys} presents the maps for the recrystallization of different film thicknesses. It is clearly evident that the crystallization threshold decreases from the 80~nm thin film to the 400~nm film, with only a small further reduction for film thickness above 400~nm. The strong thickness dependence for the thinner films can be attributed to the penetration depth of the 480~nm wavelength light, which for the thinnest films results in only partial absorption of the optical energy, which combined with heat diffusion out of the film results in a higher power threshold for crystallization\cite{Ning2022}.

For the films of 80-200~nm in thickness, a fully reversible transition to the crystalline state is achievable with pulses as short as 10~$\mu$s at a power of 12~mW. For the film thicknesses of 400~nm and above, the recrystallization is limited to quite low power levels, as indicated by the black dashed lines delineating the transition to partial crystallization and more complex dynamics. Similar to the case of amorphization, the onset of damage becomes more prominent with increasing film thickness, substantially narrowing the window for fully reversible crystallization to a fairly narrow band of pulse times and powers which curves up for shorter pulse lengths. Within the crystallization window, selecting pulses of higher power and short pulse length $< 1$~$\mu$s results in a limited transition and a partial crystalline state as is expected according to the amorphization trends seen in Fig.~\ref{fig:amo}. For the thickest films, at low powers around 7~mW and longer pulse durations of around 10~$\mu$s, a full transition back to the crystalline state can be observed. 

Together the maps for the two thickest films indicate that there is an imposed lower limit on the duration of pulses required to complete the recrystallization of the amorphous domains. In the middle band of the maps, recrystallization is also achieved but is also partial, which may be attributed to the increased pulse energies resulting in temperatures significantly above the crystallization temperature where competing processes such as melting and amorphization contribute to the dynamics. In the top right region of the maps, outlined by the green dashed lines, an abrupt reduction in the reflectance is seen, which we interpret as capping layer damage. In principle this issue could be alleviated by using more complex pulse schemes, such as the use of a series of shorter pulses to progressively recrystallize or trailing  pulses to maintain a peak temperature more suitable for crystallization. Optimal pulses required for a full transition back to a crystalline state are located in the middle of the quite narrow band of ideal crystallization conditions, and these are the pulse parameters we use for the subsequent endurance experiments. 

\subsection{Endurance of Sb$_2$Se$_3$ under fast optical cycling.}

\begin{figure}[t!]
\centering\includegraphics[width=\textwidth]{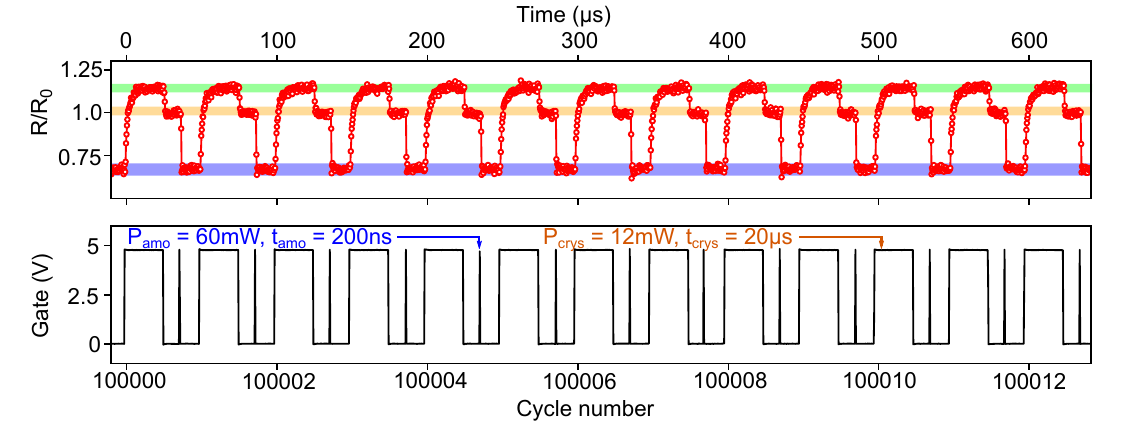} 
\caption{Example of optical modulation of the 150~nm thick film of Sb\textsubscript{2}Se\textsubscript{3} via direct optical writing after 10$^5$ cycles at 20~kHz cycling rate. Top panel: measured reflectance  for the 980~nm probe laser during the cycling between crystalline and amorphous states, normalized to the reflectance of the crystalline background (R=1.0, orange band). Blue shaded band denotes the average reflectance in the amorphous state, green band indicates the level of the thermo-optic effect during the presence of the crystallization laser. Bottom panel is the digital gate voltage applied to the laser head to gate optical pulses. Pulse parameters for the crystallization and amorphization are $P\textsubscript{crys} = 12$~mW, $t\textsubscript{crys} = 20$~$\mu$s, $P\textsubscript{amo} = 60$~mW, $t\textsubscript{amo} = 200$~ns, respectively.}
\label{fig:cyclingexample}
\end{figure}

Having fully optimized the pulse parameters for amorphization and crystallization for each film thickness, the selected optimal pulses were used in a series of optical cycling experiments. To test the optical endurance using the parameters found in the two previous experiments we performed cycling at 20~kHz for three selected samples of 80~nm, 150~nm and 800~nm thickness. In all cases we started from the thermally crystallized sample and used an amorphization pulse as the first pulse to write an amorphous area on the crystalline background. The 20~kHz frequency was found to provide sufficient time for the switched micrometer-sized region to cool to below the crystallization temperature. The relatively high cycling rate was important to allow reducing the total measurement time below the long-term drift of the setup as discussed in the Methods section.

An example of the optical modulation of the 150~nm thick Sb$_2$Se$_3$ film is shown in Fig.~\ref{fig:cyclingexample}, where we show raw reflectivity data after $10^5$ cycles at 20~kHz cycling rate. The top panel shows the reflectance $R/R_0$ normalized to the reflectance of the thermally crystallized film before switching, $R_0$. The bottom panel indicates the digital modulation trigger sequence, corresponding to the long, 20~$\mu$s crystallization pulses and short, 200~ns amorphization pulses. Additionally an analog modulation was applied to vary the average output power to result in the different levels of $P_{\rm cryst}$ and $P_{\rm amo}$. The reflectivity data show three distinct levels, as indicated by the blue, red and green shaded bands. The blue band corresponds to the amorphous state of the PCM which is achieved directly following the 200~ns pulse (typical amorphization time is 40~ns \cite{Lawson2022}). The red band corresponds to the crystalline state of the PCM achieved after the end of the crystallization pulse. The green band corresponds to the steady-state reflectivity level reached {\em during} the crystallization pulse, which is higher than the crystalline state due to the presence of a thermo-optic effect which reflects the local temperature during crystallization \cite{Nobile2023}. A similar thermo-optic effect can also be observed during the much shorter nanosecond amorphization as reported in earlier studies, corresponding to the melt phase, but is not observed here due to the much lower sampling rate during cycling \cite{Lawson2022}.

\begin{figure}[th!]
\centering\includegraphics[width=\textwidth]{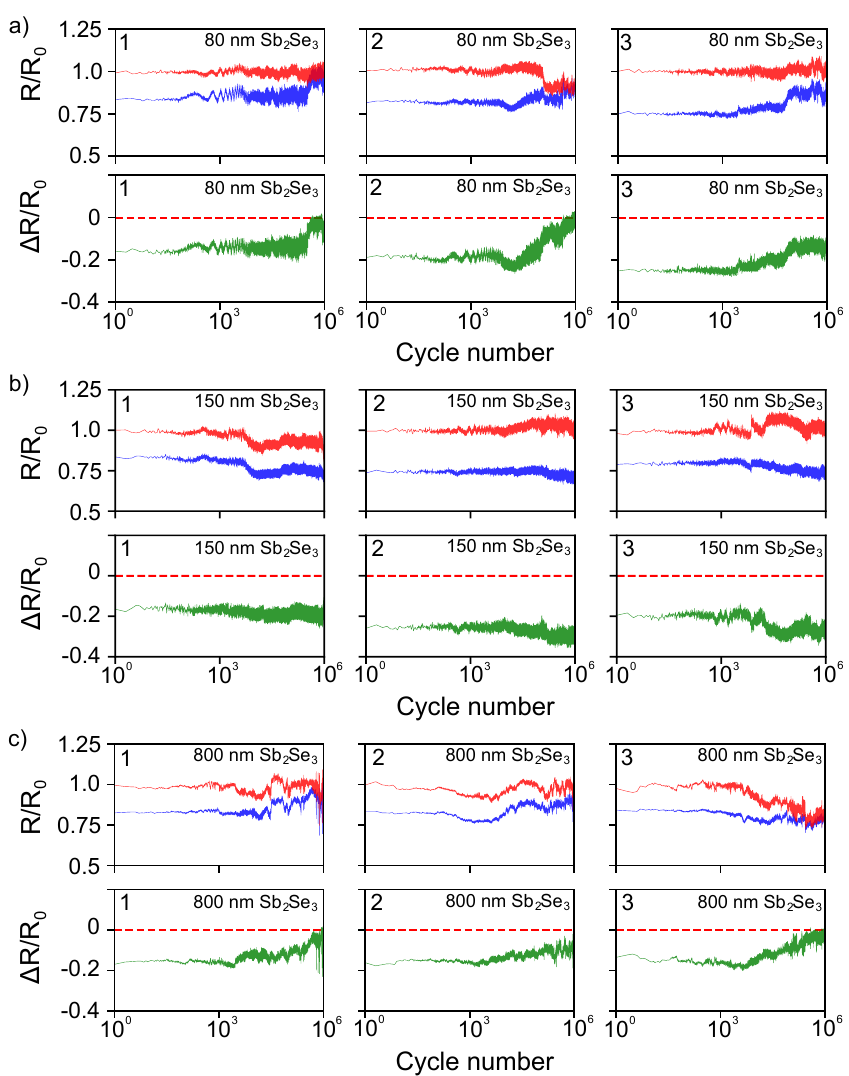}
\caption{(a-c) Endurance tests showing reflectance $R$ of the crystalline (red) and amorphous (blue) states and the associated reflectance contrast $\Delta R$ (green), normalized to initial reflectance in crystalline state $R_0$, as a function of the cycle number for Sb$_2$Se$_3$ film with thickness of 80~nm (a), 150~nm (b), and 800~nm (c). Three different endurance measurements (1-3) are shown for each film. The pulse parameters were fixed at $t_{\rm crys}=20$~$\mu$s and $P_{\rm crys}=12$~mW,  $t_{\rm amo}=200$~ns and $P_{\rm amo}= 60$~mW for (a,b) and same except for $P_{\rm crys}=7$~mW for (c) . The cycling rate was 20~kHz. Red dashed lines indicated null contrast between phases.}
\label{fig:endurance80}
\end{figure}

Typical endurance data up to $10^6$ cycles are presented in Fig.~\ref{fig:endurance80} for three samples with Sb$_2$Se$_3$ thickness of 80~nm, 150~nm, and 800~nm. The panels show the reflectivities normalized to the initial reflectivity of the crystalline layer, $R/R_0$, for the amorphous (blue) and crystalline (red) states of the PCM. The green curves show the difference $\Delta R/R_0$ between the two states. We can define the optical endurance as the number of cycles it takes for the difference in reflection between states, $\Delta R/R_0$, to fall below half the initial value. The initial contrast $\Delta R/R_0$ of around $-0.2$, is in line with the amorphization maps of Fig.~\ref{fig:amo}. Endurance values of between $10^4$ and $>10^6$ are found on different locations on the sample, which indicates some variation in the local material properties or an effect of the local substrate morphology.

\begin{figure}[th!]
\centering\includegraphics[width=0.94\textwidth]{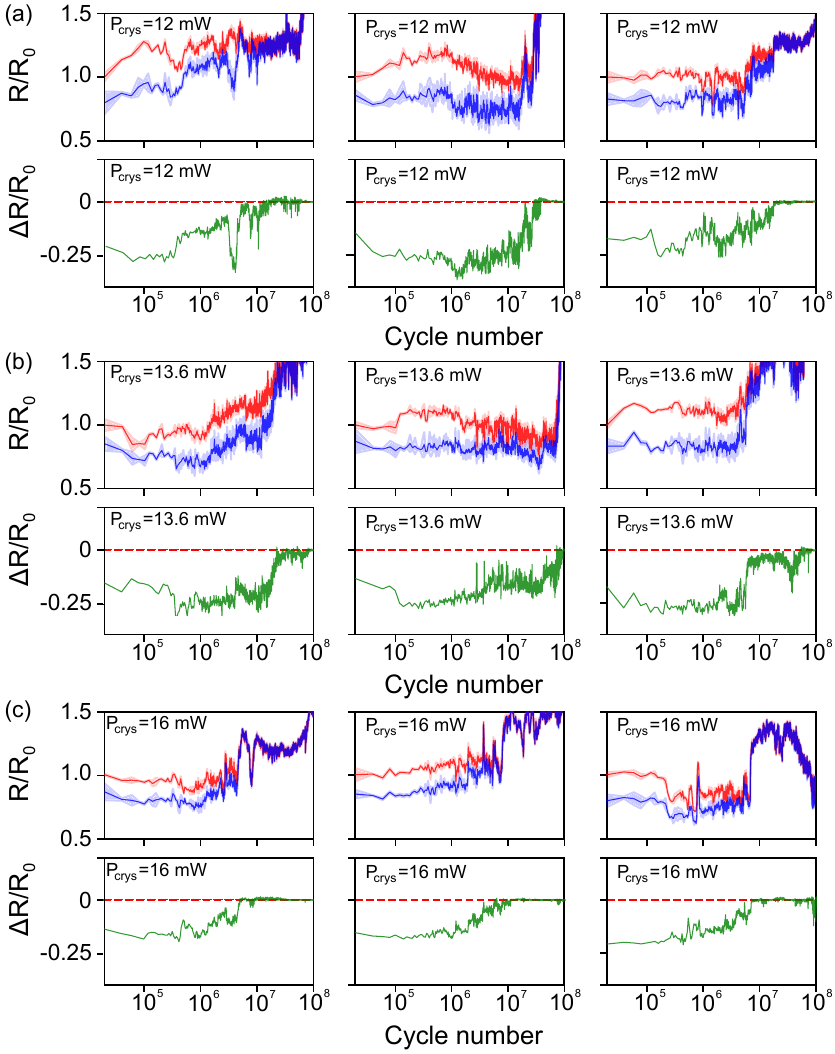}
\caption{(a-c) Endurance tests showing reflectance $R$ of the crystalline (red) and amorphous (blue) states and the associated reflectance difference $\Delta R$ (green), normalized to initial reflectance in crystalline state $R_0$, as a function of the cycle number for 150~nm thick Sb$_2$Se$_3$ film for settings of $P_{\rm crys}=12$~mW (a), 13.6~mW (b), 16.8~mW (c), for three independent experiments per setting. Other pulse parameters were $t_{\rm amo}=200$~ns, $P_{\rm amo}= 60$~mW, $t_{\rm crys}=20$~$\mu$s, cycling rate was 20~kHz. Lines indicate mean values with shaded areas the standard deviation taken over 10$^4$ cycles for each data point. Red dashed lines indicate null contrast between phases. }
\label{fig:endurancelong}
\end{figure}

The best results in our study were obtained for the 150~nm film thickness, which achieved endurance values in excess of $10^6$, with $\Delta R/R_0$ remaining within 20~\% of the initial contrast. To extend the number of cycles in the study, a different approach had to be taken to avoid excessive memory requirements associated with storing the full cycling data. In a subsequent experiment, we took statistical averages for mean and standard deviation over subsequent sets of $10^4$ cycles and as a result we were able to record results up to $10^8$ cycles. Figure~\ref{fig:endurancelong} shows results for the 150~nm sample, where we varied the crystallization power $P_{\rm crys}$ from 12 to 16.8~mW. For each power setting, we present three independent cycling results. In several runs we observe endurance of around $10^7$ cycles. These results demonstrate that an excellent durability, significantly beyond what has been demonstrated to date, can be achieved for Sb$_2$Se$_3$ films when using  optimized switching parameters. Furthermore, in many cases we obtain a full retention of the PCM contrast exceeding 10\textsuperscript{6} cycles, with little to no amorphous state drift. This in itself provides evidence that Sb$_2$Se$_3$ or nearby deviations in stoichiometries are capable of high cycling optical endurances but also strong retention of the individual states.

\section{Discussion}

In the cycling of the 80~nm and 150~nm films there seemingly exist two or more mechanisms for the failure of the Sb$_2$Se$_3$, that manifest in
the form of changes in both the crystalline and amorphous state levels. In some cases, in Fig.~\ref{fig:endurance80} and \ref{fig:endurancelong}, these changes appear independently for the different phases (i.e in Fig.~\ref{fig:endurance80} A.1 and A.2), which affirms that there is not solely a change in the complex refractive index which in turn leads to a change in the achieved reflectance for the same pulse energy. This implies not only the change in PCM optical properties but a change to the crystallization and amorphization kinetics also. We also identify a significant increase of reflectivity $R/R_0$ in both states which occurs typically above 10$^7$ cycle numbers and which could imply an increased metallicity of the layer, associated with the breakdown of the PCM switching response and loss of $\Delta R/R_0$.

In electrical switching, the change in the physical properties of the PCM states is often attributed to the change in the volume’s stoichiometry via means of drift driven by high elemental diffusivity at high temperatures in the transient phase, leading to the formation of voids and
filaments. Furthermore, higher than the crystallisation point temperatures in PCMs during thermal annealing can also lead to anisotropic crystallite formation\cite{Luong2022,Petroni2022}. In free space optical switching, atomic diffusion occurs radially due to the radial
thermal gradients induced by Gaussian laser spots. In the crystallization experiments performed by Debunne et al.\cite{Debunne2011} on uncapped stoichiometric Ge$_2$Sb$_2$Te$_5$ this results in the in-diffusion of Sb
and Te atoms to the central heated region, resulting in crystallization induced segregation. The resulting shift in the crystallization thresholds is also partially attributed
to the evaporative loss of Te and the oxidization of Ge. In our work, the evaporative losses are eliminated by the use of a capping layer and the kinetic changes are expected to arise mainly from atomic transport in the PCM itself, though failure of the capping layer also contributes to the lost of switching properties. 

The high endurance demonstrated in this work was confirmed for different regions and different samples. This highlighted a variability in the endurance performance. Whilst the maximal durability of Sb\textsubscript{2}Se\textsubscript{3} here is on the order of 10\textsuperscript{7},  in other places, although sufficiently large, the performance does not exceed 10\textsuperscript{6}. It is thought that these discrepancies are brought about due to film inhomogeneity, which acts to change the films' point-to-point switching behaviour and optical properties, the prior being the lead cause of the loss of switching. Compounded with the small windows for switching and stochastic behaviour in PCMs this imposes an important challenge in the assessment of endurance characteristics and also in the design of high durability PCM elements for integrated photonics and free-space optics. While our results indicate what might be achievable when the material works at its best, in many applications failure will be caused by the low-endurance outliers, therefore achieving consistency of performance is one of the future challenges that will determine the success of optical PCMs in applications. The required improvements could be in a passive form such as adjusting the deposition properties, further optimising stoichiometry, matching of surrounding materials, shape optimisation, but also active such as performing global re-melts to reset the material after a certain cycle number.

Such an active improvement by resetting the material is apparent in our current work where the quality of switching maps is significantly better than previous studies where an as-grown amorphous Sb$_2$Se$_3$ layer \cite{Lawson2022} was used. In particular, we see a suppression of the stochastic variations in the growth-dominated material, which is attributed to the fact that, in the current work, we recrystallize a small volume of material from a surrounding crystalline region. At the same time, we have noticed a significant, two orders of magnitude speed up of the recrystallization process as compared to the slower crystallization over millisecond time scales when starting from a previously pristine amorphous material. Slow switching of Sb$_2$Se$_3$ was attributed to bond stiffening due to the increased number of shared electrons by M\"uller et al. \cite{Muller2022} who indicated the challenge of finding low-loss optical PCMs with sufficiently fast response times. Clearly, our studies indicate that regrowth from a pre-crystallized domain can be used as a mechanism for speeding up the response time of these materials without requiring any additional material modification or doping mechanisms.

\section{Conclusions}
In conclusion, we have investigated the optical switching characteristics of thin films of the low-loss optical phase change material Sb$_2$Se$_3$ in the range of thicknesses from $80-800$~nm. We used a static tester configuration to determine the pulse parameter regimes for optically-induced amorphization and recrystallization, starting from a thermally crystallized material. These maps allowed to define the optimal pulse conditions for which reversible switching could be obtained, which were subsequently used in endurance tests. Starting from a crystallized material reduced the stochasticity compared to starting from as-grown amorphous layers, while resulting in significantly shorter crystallization times which enabled reversible switching at high cycling rates of up to 20~kHz.  The fast reversible cycling observed at up to $10^7$ cycles endurance is a significant step forward in the cycling of optical phase change materials and demonstrates that these materials, even without modifications to enhance stability, can sustain prolonged cycling. Higher speed applications might still require longer endurances. Our results pave the way for these higher endurances and already open up a significantly wider range of applications for optical PCMs in programmable photonics previously restricted by reports of low durability. Further improvements in both durability and stochasticity will be achieved with careful thermo-optical design of the system and by optimising the deposition conditions and stoichiometry of the material further. As we are expanding the limits of what this technology can offer it becomes imperative to treat the arising challenges also at a system level and provide tailored optimisation strategies for specific applications. We therefore envision further improvements to include the surrounding materials but also employ geometrically induced properties.

\begin{backmatter}
\bmsection{Funding}
The work in this paper is supported by the Engineering and Physical Sciences Research Council (EPSRC) in part through Grant EP/M015130/1, Manufacturing and Application of Next Generation Chalcogenides. DL acknowledges support through an EPSRC PhD studentship.

\bmsection{Disclosures}
The authors declare no conflicts of interest.

\bmsection{Data Availability Statement}
The data that support the findings of this study are openly available in the University of Southampton at https://doi.org/10.5258/SOTON/D2790.

\end{backmatter}



\bibliography{OME_endurance_DL}

\end{document}